\documentclass[12pt,a4paper]{article}

\usepackage[ansinew] {inputenc}
\usepackage{geometry}
\usepackage{graphicx}
\usepackage{float}
\usepackage{alltt}
\usepackage{array}
\usepackage{amsmath}

\usepackage{amssymb}
\usepackage{amsfonts}
\usepackage{amsthm}
\usepackage[hang,small,bf]{caption}
\usepackage{hyperref}
\hypersetup{colorlinks=true,linkcolor=blue}
\begin{document}
\title{A model-insensitive determination of First-hitting-time densities with Application to Equity default-swaps}
\author{\bf{Alex Langnau}\\
Allianz Investment Management, LMU Munich}

\date{}
\maketitle
\begin{abstract}
Equity default-swaps pay the holder a fixed amount of money when the underlying spot level touches a (far-down) barrier during the life of the instrument. While most pricing models give reasonable results when the barrier lies within the range of liquidly traded strikes of  plain-vanilla option prices, the situation is more involved for extremely out-of-the money barriers. In this paper we discuss a model-insensitive approach for the determination of first hitting times that does not rely on the full a priori knowledge of the stochastic process for the price dynamics. Hence more robust pricing and hedging results are expected as a result of this analysis. In contrast to stochastic volatility-models our approach is well suited for the conservative pricing of equity default-swaps.
\end{abstract}

\newpage
\pagestyle{plain}
\pagenumbering{roman}
\newpage
\tableofcontents
\newpage
\pagenumbering{arabic}

\section{Introduction}
An American Digital Put (Am-DIP) with (lower) barrier B pays the amount of $1$ EUR at maturity $T$ in case that the underlying spot level $S_t$ has touched the barrier during the life of the contract, e.g.
\begin{equation}\label{eq:1}
Am - DIP(T,S_T,B,T) = \mathbf{1}_{\tau_B<T}
\end{equation}
where $Am - DIP(t,S_t,B,T)$ denotes the price of the Am-DIP at time $t$ and $\tau_B$ denotes the stopping time of the price process $S_t$ related to the barrier $B$. Standard option pricing theory assumes the existence of an equivalent martingale measure. This implies that the fair value of payoff \eqref{eq:1} is given by an expectation under the forward adjusted probability measure $Q_T$ \footnote{Throughout this paper interest rates are assumed non-stochastic so that the forward-adjusted and risk-neutral measure are used interchangeably in the sequel},e.g.
\begin{equation}\label{eq:2}
Am - DIP(t,S_t,B,T) = Df \; E^{Q_T}(\mathbf{1}_{\tau_B<T} | \mathcal{F}_t)
\end{equation}
where, as usual, $Df$ denotes the discount factor to maturity and $\mathcal{F}_t$ is the filtration in the probability triple $(\Omega,P,\mathcal{F})$.
\\
\\
In order to evaluate \eqref{eq:2}, standard modeling procedure requires the prior knowledge of the full dynamics of the spot process $S_t$. For example, in the case of a continuously trading economy where the filtration is generated by two Brownian motions $W_t$,$\tilde{W}_t$, a common class of processes have the following form 
\begin{align}\label{eq:3}
\frac{dS_t}{S_t} &= \mu(t,\sigma_t)dt + \sigma_t dW_t \notag \\
d\sigma_t^2 &= \eta(t,\sigma_t)dt + \nu(t,\sigma_t)d\tilde{W}_t
\end{align}
where the correlation $\rho$ between the two Brownian motions is given by
\begin{equation}\label{eq:4}
<dW_t,d\tilde{W}_t> = \rho \ dt
\end{equation}
$S_t,\sigma_t,\mu(\dots),\eta(\dots),\nu(\dots)$ denote the spot, volatility, drift of the spot, drift of the volatility and vol-of-vol respectively.
\\
For example, if one sets $\eta(t,\sigma_t) = \kappa(\Theta-\sigma_t^2)$ and $\nu(t,\sigma_t) = \sqrt{\sigma_t^2}w$ for some level of the mean-reversion speed $\kappa$ and mean-reversion level $\Theta$ as well as a vol-of-vol parameter $w$, one obtains the Heston model of stochastic volatility in this case. Note that the dynamics defined in \eqref{eq:3} does not necessarily lead to an equivalent martingale measure. Without specifying the conditions in detail, for the purpose of this paper, we restrict the functions defined in \eqref{eq:3} such that the Novikov condition is fulfilled for the variance process at all times.
\\
\\
As far as the pricing of Am-DIP is concerned, \eqref{eq:3},\eqref{eq:4} can readily be integrated to obtain  the expectation value of \eqref{eq:2}. However even after a successful completion of this task the following questions remain:

\renewcommand{\labelenumi}{\roman{enumi}}
\begin{enumerate}
\item
How does the pricing of the American-DIP depend on the particular choice of the dynamics in \eqref{eq:3}?  If there were to exist two different models that both are consistent with the market prices for plain-vanilla options, which model should one prefer in practice and how can one specify the model-sensitivity more generally for the Am-DIP?
\item
How can one establish confidence in the parameterization of \eqref{eq:3} in situations where the barrier level B is below the traded strike range of the plain vanilla option market and hence no direct calibration of the parameters is possible?
\item
How can a conservative price be constructed in the pricing framework of \eqref{eq:3} and how can the hedges be best constructed and interpreted?
\end{enumerate}

In this paper we pursue an alternative pricing methodology for the American digital put option. Our approach utilizes a well-known intuitive relationship between American and European digital puts options in the Black-Scholes world and generalizes the latter to a situation where a volatility skew is present.
\\
\\
Hence our approach is primarily driven by hedging intuition of the American digital put and is, in this sense, conceptually different to the approach outlined above. In this paper we argue that the full knowledge of the dynamics is not required in order to solve the pricing problem of \eqref{eq:2}.  Hence our approach relies on fewer specific assumptions and can be viewed as a framework that simultaneously describes a wide range of different stochastic processes. This is in contrast to the standard option pricing philosophy of \eqref{eq:3} which requires the full \textit{a-priori} knowledge of the stochastic process.
\\
\\
Note that the idea of relating the valuation of knock-out options directly to the hedge is not a new idea. Under the special assumptions of driftless spot processes and symmetric distributions Carr et al \cite{Carr} describe how barrier options can be written as a portfolio of plain-vanilla options. In a later paper Carr and Lee generalize Put-Call Symmetry to a wider class of processes \cite{Carr2}. However two different pricing models that are both consistent with the plain-vanilla market will generally yield different values for knock-out options if these assumptions are relaxed. Hence knock-outs exhibit "`exotic risk"' that cannot be hedged into plain-vanilla instruments. As will be discussed in the paper such valuation differences can arise due to different dividend assumptions. In addition forward distributions for the spot are not uniquely determined by the plain-vanilla space either. As will be discussed in a later chapter this will lead to different unwinding costs of the hedge at the barrier which in turn will affect the price of the American digital put and hence give rise to model-risk.
\\
\\
The paper is outlined as follows:
\\
\\
Chapter 2 recaps a well-known approximate  relationship between American and digital puts in the case of a Black-Scholes world.
\\
\\
Chapter 3 derives an integral-equation for the stopping time density in the presence of a volatility skew.
\\
\\
Chapter 4 compares a specific instantaneous pricing model to our approach. We show how prices can be reconciled with our methodology once the forward volatility skew as well as the dividend assumptions are known at the barrier. Hence as a result one can think of our approach as a "pricing framework" that simultaneous comprises a large class of different pricing models.
\\
\\
Chapter 5 discusses the skew sensitivity of an American Digital puts.
\\
\\
Chapter 6 discusses an example for the conservative pricing of an equity default swap.

\pagebreak
\section{A "handy" relationship between American Digital Puts and European Digital Puts in a Black-Scholes world}
In contrast to an Am-DIP, European digital puts (Eur-DIP) pay an amount of 1 Euro in the case that the terminal spot price is below a barrier $B$ at the time of maturity, e.g
\begin{equation}\label{eq:5}
Eur-DIP(T,S_T,B,T) = \mathbf{1}_{S_T<B}
\end{equation}
where $Eur-DIP(t,S_t,B,T)$ denotes the price of a Eur-DIP at time $t$ with strike $B$. Hence
\[
Eur-DIP(t,S_t,B,T) = Df \; E^{Q_T}(\mathbf{1}_{S_T<B} | \mathcal{F}_t)
\]
Hence Eur-DIP are "non-exotic" as its price is a direct consequence of the plain-vanilla European option market. The relationship is given by
\begin{equation}\label{eq:7}
Eur-DIP(t,S_t,B,T) = \frac{dPut(t,T,K=B)}{dK}
\end{equation}
where $Put(t,T,K)$ denotes the value of a plain-vanilla European put of strike $K$ and is given by the Black-Scholes equation
\begin{equation}\label{eq:8}
Put(t,T,K) = Df \; (K \mathcal{N}(-d_2) - F \mathcal{N}(-d_1))
\end{equation}
where
\[
d_{1/2} = \frac{ln(\frac{F}{K})\pm \frac{1}{2} \sigma^2 (T-t)}{\sigma\sqrt{T-t}}
\]
In the particular case of a Black-Scholes world, the volatility does not depend on the strike and the rhs of \eqref{eq:7} simplifies to
\begin{equation}\label{eq:10}
\frac{dPut(t,T,K=B)}{dK} = Df \; \mathcal{N}(-d_2)
\end{equation}
Here $F,B,\sigma,T$ denote the forward, strike, volatility and maturity respectively.
\\
\\
Eq. \eqref{eq:7} states that the price of Eur-DIP is model-independent, given the derivative of the European Puts with respect to the strike.
\\
\\
Eq. \eqref{eq:10} allows one to derive an approximate relationship between the Am-DIP and the  Eur-DIP in a Black-Scholes world. In order to see this we note that at the time $\tau_B$ of hitting the barrier $B$, the forward price and the barrier price is approximately the same, e.g
\[
F_{\tau_B}(T) = S_{\tau_B}(1+\Delta) = B(1+\Delta)
\]
with a small correction $\Delta$. Hence the value of a Eur-DIP at the barrier is given by
\[
Eur-DIP(\tau_B,B,B,T-\tau_B) = Df \; \mathcal{N}(-\frac{log(1+\Delta)}{\sigma\sqrt{T-\tau_B}} + \frac{1}{2}\sigma\sqrt{T-\tau_B})
\]
In most cases one observes for sufficiently small values for $T-\tau_B $ that $-\frac{log(1+\Delta)}{\sigma\sqrt{T-\tau_B}}+\frac{1}{2}\sigma\sqrt{T-\tau_B} \approx 0$ and hence
\begin{equation}\label{eq:13}
Eur-DIP(\tau_B,B,B,T-\tau_B) \approx Df \; \frac{1}{2}
\end{equation}
On the other hand, at the time of hitting the barrier, the price of the Am-DIP put equals the discount factor to maturity so that the following approximate relationship is obtained
\begin{equation}\label{eq:14}
Am-DIP(t,S_t,B,T) \approx 2 \; Eur-DIP(t,S_t,B,T)
\end{equation}
In order to interpret \eqref{eq:14} intuitively, assume that one goes short an American Digital Put of notional $N$ at time $0$. As a hedge \eqref{eq:14} suggests to go long a European Digital Put option of notional $2N$ at the same time.
\\
\\
The following two scenarios are of relevance:

\renewcommand{\labelenumi}{\arabic{enumi}}
\begin{enumerate}
\item
The spot never touches the barrier prior to maturity
\\
\\
In this case neither the American nor the European Digital Put are paying out.
Hence one breaks even in this case.
\item
The spot touches the barrier prior to maturity
\\
\\
The American Put is paying out the notional $N$ at maturity in this case. The
European-Digital-Put on the other hand is at the money when the barrier is
touched. This implies that the hedge is worth approximately $50\%$ of the notional
$2N$ at that time (see \eqref{eq:13}). Hence the money made by liquidation of the hedge at 
the time of touching the barrier finances the Am-DIP position in this case. 
\end{enumerate}
Hence the hedging argument presented above intuitively "explains" \eqref{eq:14}.
\\
\\
However, the precise relationship between Am-DIP and Eur-DIP depends on the exact value of $\mathcal{N}(-\frac{log(1+\Delta)}{\sigma\sqrt{T-\tau_B}}+\frac{1}{2}\sigma\sqrt{T-\tau_B})$ which does not equal $\frac{1}{2}$ even in a Black-Scholes world.
\\
\\
In reality also the price of a forward contract at time $\tau_B$ is of relevance in addition to the level of Black-Scholes volatility at that time.
\\
\\
In addition, when a volatility skew is present, Eq. \eqref{eq:7} implies an additional vega term such that  \eqref{eq:7} can be  written as
\begin{equation}\label{eq:15}
Eur-DIP(t,S_t,B,T) = Df \ \mathcal{N}(-d_2)+vega \; \sigma'(B)
\end{equation}
where $\sigma'(B)$ denotes the slope of the Black-Scholes implied volatility at the barrier $B$. Eq.\eqref{eq:15} also states that the price of the Eur-DIP decreases with the volatility-skew increasing. This observation will prove important in later sections.
\\
\\
The vega term in \eqref{eq:15} describes the contribution of the volatility skew to the price of the Eur-DIP. It becomes particularly important when the barrier is hit and the Eur-DIP hedge must be unwound. The vega term affects the price of the portfolio at that time and quantifies the price impact that is due to the deviation from log-normality. It is instructive to attach some numbers to a specific example in order to get a better understanding for the impact of the volatility skew.
\\
\\
Consider a one month European Put spread struck at the money. The European Put spread prices at $50.5\%$ if the vega term is ignored in \eqref{eq:15}. The correct value that includes the vega contribution is $44.7\%$. Hence on a notional of 800 Mio USD the pricing difference will be 46 Mio USD.  
\\
\\
Hence one concludes that, in the presence of a forward-volatility skew, an additional deviation from the "unwind-factor" $\frac{1}{2}$ occurs that is due to the at-the-money slope of the forward volatility skew at the time of hitting.
\pagebreak
\section{Pricing methodology}
The valuation of a Am-DIP requires the knowledge of the stopping time density $\rho(\tau)d\tau$ which is the probability of first hitting the barrier between the time $\tau$ and $\tau+d\tau$. 
\\
\\
However, the knowledge of $\rho(\tau)d\tau$ also allows to relate Eur-DIP at different valuation times: From
\begin{align*}
Eur-DIP(0,S_0,B,T) &=  E^\beta[\mathbf{1}_{S_T<B}| \mathcal{F}_0] =  E^\beta[\mathbf{1}_{S_T<B}\mathbf{1}_{\tau_B<T}| \mathcal{F}_0] \\
&= \int_0^Td\tau \; E^\beta[\mathbf{1}_{\tau_B \in d\tau}| \mathcal{F}_0]E^\beta[\mathbf{1}_{S_T<B}| \tau_B \in d\tau,  \mathcal{F}_0]
\end{align*}
where $E^\beta[\dots]$ denotes the expectation under the risk-neutral measure. Without loss of generality \footnote{Note that throughout this paper interest rates are assumed non-stochastic }we have set the discount factors to one in order not to unnecessarily complicate the equations. 
It follows that
\begin{equation}\label{eq:16}
Eur-DIP(0,S_0,B,T) = \int_0^Td\tau \rho(\tau) \; EurDIP(F_\tau(T),B,T-\tau,\hat{\sigma}_{B,\tau,T-\tau},\hat{\kappa}_{B,t,T-\tau})
\end{equation}
\\
The variables $\hat{\sigma}_{B,\tau,T-\tau}$ describe the "forward - volatility" at the stopping time $\tau$ of an option with maturity $T$ when the spot price is first at the level $B$. Similarly $\hat{\kappa}_{B,\tau,T-\tau}$ describes the "forward-at-the-money-volatility-slope" at time $\tau$.
\\
Eq. \eqref{eq:16} defines the stopping time density
\[
d\tau \; \rho(\tau) \equiv E^\beta[\mathbf{1}_{\tau_B \in d\tau} | \mathcal{F}_0]
\]
and
\begin{equation}\label{eq:17}
EurDIP(F_\tau(T),B,T-\tau,\hat{\sigma}_{B,\tau,T-\tau},\hat{\kappa}_{B,\tau,T-\tau}) \equiv E^\beta[\mathbf{1}_{S_T<B}|\tau_B \in d\tau,  \mathcal{F}_0]
\end{equation}
\\
\\
where the explicit form of the lhs of \eqref{eq:17} follows  from \eqref{eq:15} and \eqref{eq:8}.
\\
\\
Eq.\eqref{eq:16} defines an integral equation for the unknown stopping time $\rho(\tau)$. It will play a crucial role in our methodology.
\\
\\
For a given complete set of plain vanilla option prices, the lhs of \eqref{eq:16} is fully specified. On the rhs, the value of  $Eur-DIP(\dots)$  only depends on the forward volatility $\hat{\sigma}_{B,\tau,T-\tau}$, and the forward-at-the-money volatility slope $\hat{\kappa}_{B,\tau,T-\tau}$ in conjunction with the forward price $F_\tau(T)$. In the case where these quantities are specified Eq.\eqref{eq:16} can be solved for $\rho(\tau) $ without the explicit knowledge of the stochastic process. This is the strategy followed in this paper.
\\
\\
Putting it another way, two different pricing models of the type outlined in \eqref{eq:3} that reproduce identical prices for the vanilla options, have different  stopping time densities if at least one of the following statements is correct:
\\
\\
\renewcommand{\labelenumi}{\roman{enumi}}
\begin{enumerate}
\item
The models disagree on the level of forward volatility of an at-the-money call at the time when the barrier is hit $\hat{\sigma}_{B,\tau,T-\tau}$.
\item
The at-the-money slope of the forward volatility surface is different at the time of hitting $\hat{\kappa}_{B,\tau,T-\tau}$.
\item
The forward price $F_\tau(T)$ is different at the time of touching the barriers.
\end{enumerate}
Note that case (iii) typically occurs when two models have different dividend assumptions despite the fact that $F_0(T)$ is the same in both cases: A proportional dividend model will generally exhibit a higher value for $F_\tau(T); \; \tau > 0$ compared to a model where dividend amounts are assumed to be independent of spot.
\\
\\
From this discussion it follows that, for the purpose of  pricing Am-DIPs, details of the stochastic process only matter to the extent that they produce differences in at least one of the quantities above. Should continuous pricing models produce the same values for $\hat{\sigma}_{B,\tau,T-\tau}$,$\hat{\kappa}_{B,\tau,T-\tau}$,$F_\tau(T)$ in addition to identical plain vanilla market prices, they also agree on the price of Am-DIP as well.
\\
\\
In order to solve \eqref{eq:16} we discretise the equation according to 
\[
T_n = n\Delta T
\]
where $T_N = T$ and $T_0 = 0$ with the result
\begin{align}\label{eq:19}
&EurDIP(S_0,B,T_{n+1},\hat{\sigma}_{B,0,T_{n+1}},\hat{\kappa}_{B,0,T_{n+1}}) \notag \\
&=\int_0^{T_n}d\tau \rho(\tau) \; EurDIP(B,B,T_n-\tau,\hat{\sigma}_{B,\tau,T_n-\tau},\hat{\kappa}_{B,\tau,T_n-\tau}) \notag \\
&+ \Delta T \rho(T_n) \; EurDIP(B,B,\Delta T,\hat{\sigma}_{B,T_n,\Delta T},\hat{\kappa}_{B,T_n,\Delta T})
\end{align}
for $n=0,\dots,N-1$.
\\
\\
Eq. \eqref{eq:19} states the main result of this paper. It defines a recursive equation for $\rho(.)$ which can easily be solved numerically provided that the quantities outlined in (i),(ii),(iii) are known.
\\
\\
Note that \eqref{eq:19} relates the stopping time density directly to Eur-DIP-prices as well as the forward-vol and forward-vol-slope at the time of hitting. In the case where these values are provided, \eqref{eq:19} allows the calculation of $\rho(.)$ without the explicit prior knowledge of the full stochastic process for $S_t$. Even though the quantities $\hat{\sigma}_{B,\tau,T-\tau}$,$\hat{\kappa}_{B,\tau,T-\tau}$  are not explicitly traded, they represent "market-variables" \footnote{as opposed to model parameters}. Hence a trader should have good intuition about these values together with (conservative) bounds for them.
\\
\\
In the remaining chapters of this paper we will apply \eqref{eq:19} explicitly and discuss consequences.
\\
\\
Once the stopping time densities have been found, the American digital put can be valued according to 
\begin{align}\label{eq:19a}
Am-DIP(B) = Df(T) \; \int_0^T d\tau\rho(\tau)
\end{align}
In cases where the payment is made at the time of hitting the price is given by
\[
Am-DIP(B) = \int_0^T d\tau \rho(\tau) Df(\tau)
\]
where $Df(t)$ denotes the discount factor at time $t$.
\pagebreak
\section{Relationship with standard Pricing models}
Eq.\eqref{eq:19} and Eq.\eqref{eq:19a} state that the stopping time densities can be determined if the cost of unwinding the hedge can be calculated. This is the case if the level and the slope of volatility is known at the time the barrier is first hit. This means that our pricing methodology in effect describes a whole class of specific diffusion models simultaneously.
\\
\\
In order to demonstrate this more explicitly let us price a one year Am-DIP on STOXX50E struck at $90\%$ of spot.
\\
\\
The Black Scholes Barrier Price is $64\%$ whereas the local vol price turns out to be $54.9\%$. 
\\
\\
If one pre-calculates at-the-barrier-forward-volatilities together with its slope in the local-volatility model and subsequently uses these values in Eq.\eqref{eq:19} we obtain a value of $54.8\%$ in our approach.
\\
\\
Hence our pricing methodology coincides with local vol if one chooses the barrier vol as well as barrier vol slope obtained in this model.
\\
\\
The crux of our pricing methodology is the observation that different models can give rise to pricing differences only to the extent that their volatility slope or dividend treatment differ upon touching the barrier.
\\
\\
This offers a new way to stress the assumptions of a given model by bumping the vol and slope at the barrier calculated by the model.
\\
\\
It is well known that the local vol model underestimates the level of forward skew substantially. Our approach on the other hand allows the tuning of future skews explicitly in a realistic way.
\\
\\
\pagebreak
\section{Skew Risk in American Digital Puts}
It is very instructive to discuss some details of the skew risk in the American DIP explicitly.
\\
\\
In the following we graph the price of  a 6 month Am-DIP struck at $90\%$ of spot as a function of the barrier-skew at inception (spot-skew). The skew factor denotes the skew in units of the market volatility slope at the time of pricing. It shows that the price of an Am-DIP decreases with a steepening of the volatility skew. 
\\
\begin{figure}[hptb]
\centering
\includegraphics[width=\textwidth,height=0.4\textheight,keepaspectratio=false]{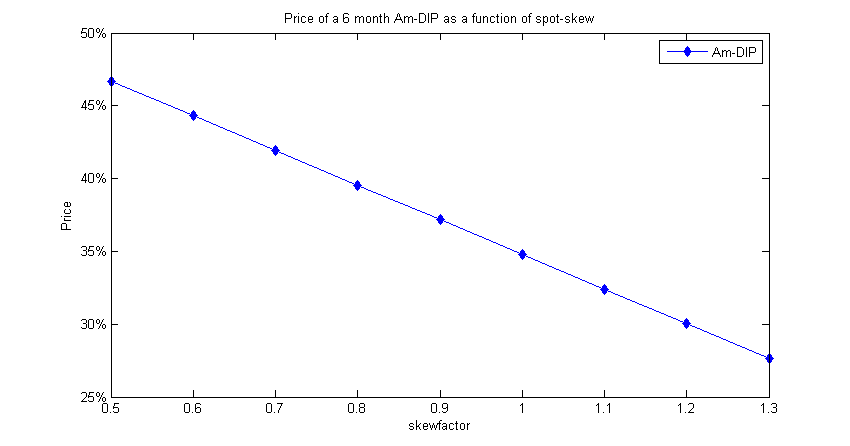}
\caption{Price of an Am-DIP as a function of spot-skew. A "`skewfactor"' of 1 corresponds to the current market level of the vol-slope. The graph shows that Am-DIPs are short spot-skew. }
\label{fig:1}
\end{figure}
\\
As outlined in the introduction Am-DIP options can be approximately replicated by a EUR-DIP position with twice the notional of the Am-DIP (see \eqref{eq:14}). Hence the flatter the spot skew the more expensive the costs of putting on the initial hedge (see also  Eq. \eqref{eq:15}) and therefore the higher the fair value of the Am-DIP. This is reflected in Fig.\ref{fig:1}. 

\begin{figure}[hptb]
\centering
\includegraphics[width=\textwidth,height=0.4\textheight,keepaspectratio=false]{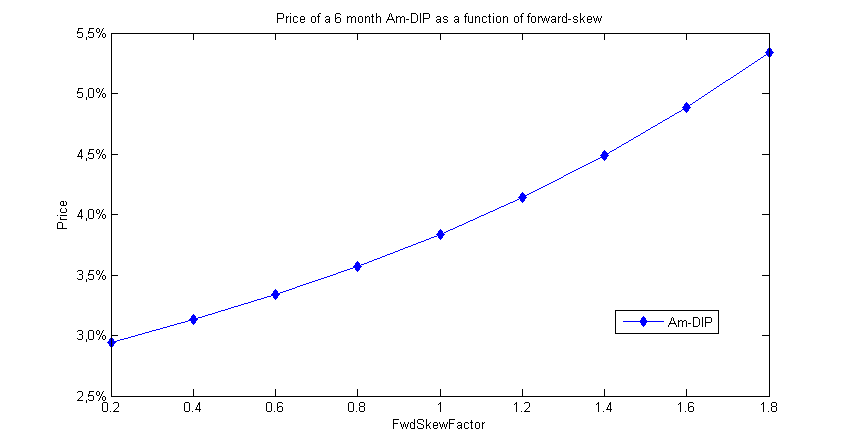}
\caption{Am-DIP sensitivity to the forward skew at the time of first hitting. A FwdSkewFactor of 1 corresponds to the spot-skew. The graph shows that Am-DIPs are long forward skew}
\label{fig:2}
\end{figure}

Graph Fig.\ref{fig:2} displays the dependence on the forward skew for a 6 month Am-DIP with barrier of $60\%$:
\\
\\
One observes that, in contrast to the short spot-skew behavior, Am-DIP are long forward-skew instead. Intuitively this is obvious: the steeper the volatility-skew at the time of unwinding the hedge the less money is retrieved which results in a higher price of the AM-DIP in this case.

\begin{figure}[hptb]
\centering
\includegraphics[width=\textwidth,height=0.4\textheight,keepaspectratio=false]{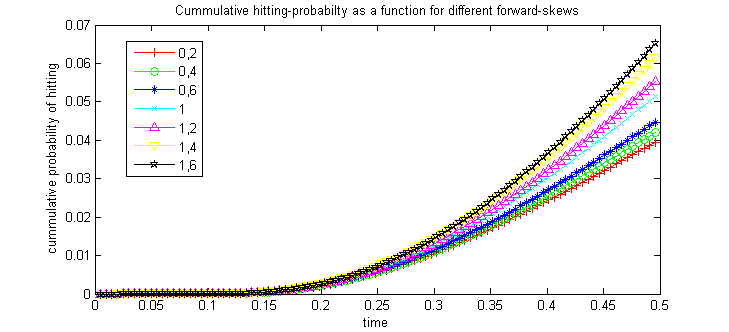}
\caption{Cumulative hitting probability for different assumption of the forward-skew as a function of time.}
\label{fig:3}
\end{figure}

Hence we summarize that the price of an AM-DIP is short spot-skew but long forward-skew at the barrier. It is questionable whether, for the purpose of determining a conservative price for the Am-DIP, the stochastic class of models outlined in \eqref{eq:3} allows one to "`tune"' spot-skew and forward-volatility skew in opposite direction.
This is in contrast to our approach where spot-skew, forward-skew and forward vol can be tuned independently such that a conservative price can be obtained.
\\
\\
\\
Fig.\ref{fig:3} shows the result of an explicit calculation of the hitting time using Eq.\eqref{eq:19} for different assumptions of the forward-skew. The steeper the forward-skew, the higher the price of the Am-DIP and hence the higher the cumulative hitting probability that is obtained from \eqref{eq:19}.
\\
\\
\pagebreak
\section{A pricing example for an Equity-default swap}

In this section we discuss the pricing of a 6 month Equity default swap (EDS) on Commerzbank with a barrier of 70\% down as of the 15 Dec 2009.
As the share price at the time traded at 5.945 Euros this implies that the Equity default swap pays out if at any time between 15-Dec-2009 and 15-Jun-2010 the price drops
below 1.78 Euros. This price drop of 70\% may not result into default necessarily but would most likely result in serious distress of the company.
The at-the-money vol traded roughly at 52\% whereas our estimates for the vol at the barrier is about 80\%. The spot barrier slope could be around -11\% according to our estimates. If one were to assume the vol and slope upon touching to be the same one obtains a price of 165 basis points for the price of the EDS. However it is impossible to obtain plain-vanilla option quotes at strikes near the barrier level.  As one is short spot skew, for the purpose of obtaining a conservative price estimate, we take 80\% of our best estimate for this quantity and obtain a value of 180 basis points instead. In the case of the forward volatility skew we increase the original value by a factor of 2 in order to obtain a new estimate of 187 basis points. Similarly one can correct the price further if one bumps the best estimate for the spot barrier-vol up further to 86\% while bumping the forward vol at the barrier down by 5\%. In this case we obtain the conservative price of 281 basis-points. In a similar way different dividend assumptions
can be tested upon hitting the barrier in order to incorporate conservative dividend assumptions into the pricing. Note that even if the pricing was done more aggressively our approach is ideally suited for calculating reserves for this trade. Finally we remark that due to liquidity issues
as well as jump risk it may not be possible to unwind the trade at exactly the barrier level. An additional shifting of the barrier towards higher levels could take these effects into account.

\section{Acknowledgement}

I would like to thank Syed Aqeel and Alexander Kabanov for constructive comments on this paper.

\bibliographystyle{plain}


\end{document}